\documentclass[review]{elsarticle}

\usepackage{lineno,hyperref}

\nolinenumbers
\usepackage{subfig}
\usepackage{url}

\usepackage[export]{adjustbox}

\journal{NIM-A}

\begin{document}
\nolinenumbers
\begin{frontmatter}

\title{A Neural Network approach to reconstructing SuperKEKB beam parameters from beamstrahlung.}

\author[cern]{S. Di Carlo\fnref{myfootnote1}}

\fntext[myfootnote1]{salvatore.di.carlo@cern.ch}

\author[wayne]{G. Bonvicini\fnref{myfootnote2}}

\fntext[myfootnote2]{gbonvicini@wayne.edu}

\author[jouf]{N.A. Althubiti}
\author[tabuk]{R.Ayad}
\author[mexicoci]{ E. De La Cruz-Burelo}
\author[sinaloa]{I. Dom\'inguez}
\author[tabuk]{B.O. El Bashir}
\author[wayne]{H. Farhat}
\author[kek]{J. Flanagan}
\author[wayne]{R. Gillard}
\author[wayne]{S. Izaguirre Gamez}
\author[kek]{K. Kanazawa}
\author[wayne]{K.Kumara}
\author[wayne]{D. Liventsev}
\author[sinaloa]{P.L.M. Podesta-Lerma}
\author[wayne]{D. Ricalde-Herrmann}
\author[sinaloa]{D. Rodriguez Perez}
\author[puebla]{G. Tejeda-Mu\~noz}
\author[kek]{M. Tobiyama}
\author[mexicoci,mexicoca]{I. Heredia de la Cruz }
\address[cern]{European Organization for Nuclear Research (CERN), 1211 Geneva, Switzerland}
\address[wayne]{Department of Physics and Astronomy - Wayne State University, 666 West Hancock, Detroit, MI 48201, USA}
\address[jouf]{Physics Department, College of Science, Jouf University, Sakaka 2014, KSA}
\address[tabuk]{Department of Physics, Faculty of Science, Tabuk University, Tabuk, 71451 Saudi Arabia}
\address[mexicoci]{Centro de Investigacion y de Estudios Avanzados (CINVESTAV), Mexico City , 07360, Mexico}
\address[sinaloa]{Universidad Autonoma de Sinaloa, Sinaloa 80000, Mexico}
\address[kek]{High Energy Accelerator Research Organization (KEK), Tsukuba 305-0801 Japan}
\address[mexicoca]{Consejo Nacional de Ciencia y Tecnologia, Mexico City, 03940, Mexico}
\address[puebla]{Benemerita Universidad Autonoma de Puebla, Puebla, Mexico}




\begin{abstract}
This work shows how it is possible to reconstruct SuperKEKB's beam parameters using a Neural Network with beamstrahlung signal from the Large Angle Beamstrahlung Monitor (LABM) as input. We describe the device, the model, and discuss the results.
\end{abstract}

\begin{keyword}
\texttt{Beamstrahlung}\sep 
\texttt{SuperKEKB}\sep 
\texttt{Collider}\sep
\texttt{Beam}\sep 
\texttt{Monitoring}\sep 
\texttt{LABM}\sep 
\texttt{Machine-Learning}\sep 
\texttt{Neural-Network}\sep
\end{keyword}

\end{frontmatter}

\nolinenumbers

\section{Introduction}
\label{section:introduction}
SuperKEKB~\cite{Ohnishi:2013fma,Akai:2018mbz} is an intersecting double storage ring particle accelerator with a circumference of 3.016 km that collides electrons and positrons to provide luminosity for the Belle II experiment~\cite{Abe:2010gxa}. SuperKEKB is the successor of KEKB~\cite{KEKBReport}. While KEKB achieved a maximum peak luminosity of $2.11\times{10}^{34}{cm}^{-2}{s}^{-1}$~\cite{Abe:2010gxa}, SuperKEKB will attempt to reach a peak luminosity of $8\times{10}^{35}{cm}^{-2}{s}^{-1}$~\cite{Ohnishi:2013fma}, about 40 times larger than its predecessor. Since the luminosity is strongly dependent on beam-optical parameters at the IP, it is important to have direct measurement of such parameters. Such information is especially crucial to a nano-beam collider as SuperKEKB, where there is strong sensitivity to small parameter changes, in order to maximize the luminosity extracted for bunch crossing. The Large Angle Beamstrahlung Monitor (LABM) is a device designed to measure the beamstrahlung emitted at the Interaction Point (IP) of a e+e- collider. Beamstrahlung is the radiation emitted by two beams of charged particles due to their electromagnetic interaction~\cite{Augustin:1978ah}. Beamstrahlung is directly related to the size and configuration of the beams, and provides direct information on the beams at the IP. Specifically, vertically unequal beams create an excess of y-polarized light as seen by the telescope observing the fatter beam (conversely, y-polarized light from the smaller beam will decrease).
\par
The latest version of the LABM is installed around the IP of SuperKEKB. The LABM at SuperKEKB measures 32 independent values, with different optical properties, that are directly related to the size and position of the beams. In this framework, the LABM can be extremely useful to monitor the beams and correct them in case they show an unwanted behavior that can cause luminosity degradation. One of the challenges of the LABM is to relate these 32 measurements to observables of interest. This can be done on theoretical grounds, by constructing a variable that is function of all or some of the 32 measurements, by traditional fitting methods, e.g., a linear regression, or using machine learning techniques, e.g. a neural network. In this paper, we will present and compare results from linear regression and neural network models. This will be an experimental validation for both the LABM and, more in general, for machine learning models applied to the first particle accelerator using the nano-beam scheme. The average beam parameters at the IP of SuperKEKB IP for the data used in this paper are given in Table~\ref{table:1}. 
\begin{table}[ht]
\centering
\begin{center}
\begin{tabular}{|c|c|c|}
\hline
$ Beam $                      & LER ($e^{+}$) & HER ($e^{-}$) \\ \hline
$ L(cm^{-2} s^{-1}) $  & \multicolumn{2}{c|}{2.8 $\times10^{34}$} \\ \hline
$ E(GeV) $                    & 4    & 7       \\ \hline
$ N (10^{10}) $ & 3.8   & 3.1       \\ \hline
$\beta^{*}_{x}(m)$ & 0.08     & 0.06       \\ \hline
$\beta^{*}_{y}(m)$ & 0.001    & 0.001     \\ \hline
$\varepsilon_{x}(nm)$ & 3.5   & 4.7   \\ \hline
$\varepsilon_{y}(pm)$     & 46.4    & 37.7      \\ \hline
$ \sigma^{*}_{x} (\mu m)$            & 16.7  & 16.7    \\ \hline
$\sigma^{*}_{y} (nm)$            & 214.8   & 192.7     \\ \hline
$\sigma^{*}_{z} (mm)$             & 6    & 5      \\ \hline
\end{tabular}
\end{center}
\caption{SuperKEKB average beam parameters at the IP for the data used in this paper. The Low Energy Ring (LER) is the positron ring, The High Energy Ring (HER) is the electron ring. L is the luminosity, E is the energy of the beam, N the number of particles per bunch, $\beta^{*}$ the $\beta$ function at the IP, $\varepsilon$ the emittance, and $\sigma^{*}$ the size of the beam at the IP.}
\label{table:1}
\end{table}
\par
This article is organized as follows. In Section~\ref{section:labm}, we describe the LABM as it was installed at SuperKEKB. In Section~\ref{section:datas}, we describe how data for this work was selected. In Section~\ref{section:nn_model}, we give an overview of the machine learning model used in this work, which is a deep Neural Network. In Section~\ref{section:beam_parameters}, we present the result of our Neural Network model and compare it with a traditional Linear Regression. In Section~\ref{section:discussion}, we discuss the results in details and provide some comments. Finally, in Section~\ref{section:conclusion}, we summarize the results presented the paper.

\section{LABM at SuperKEKB}
\label{section:labm}
\begin{figure}%
    \centering
    \subfloat[]{{\includegraphics[width=5.5cm,valign=c]{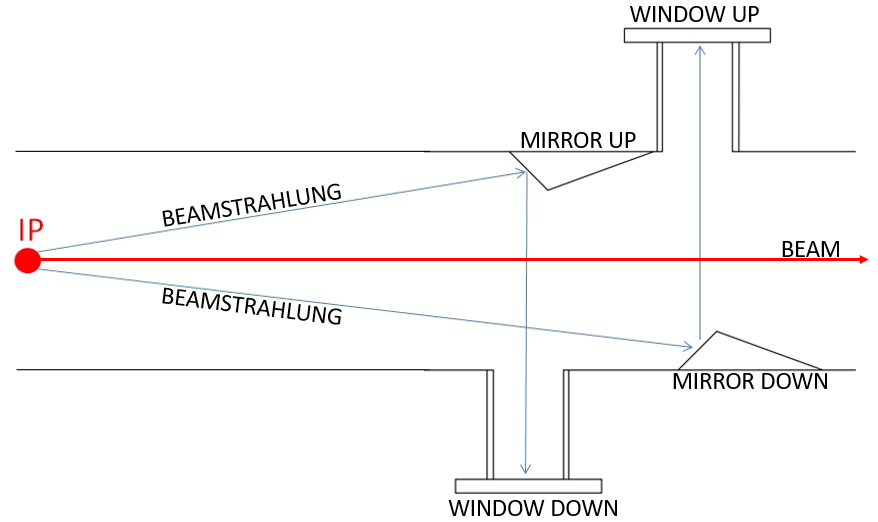} }}%
    \qquad
    \subfloat[]{{\includegraphics[width=5.5cm,valign=c]{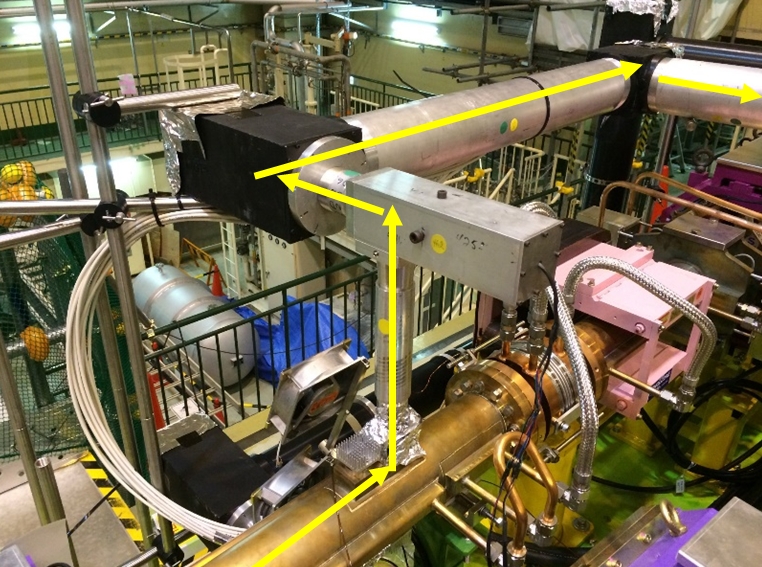} }}%
    \qquad
    \subfloat[]{{\includegraphics[width=5.5cm,valign=c]{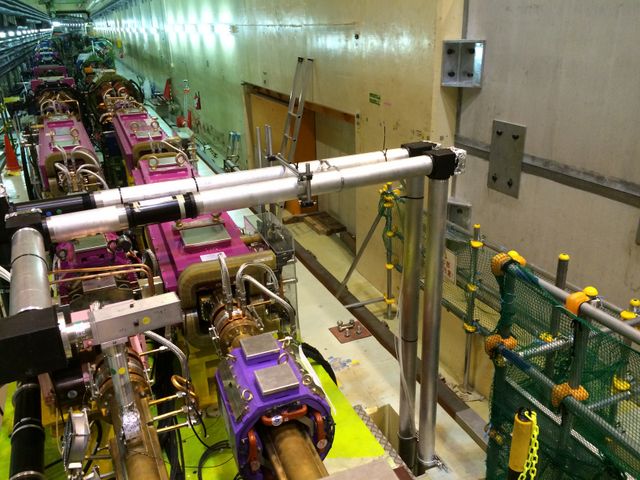} }}%
    \qquad
    \subfloat[]{{\includegraphics[width=5.5cm,valign=c]{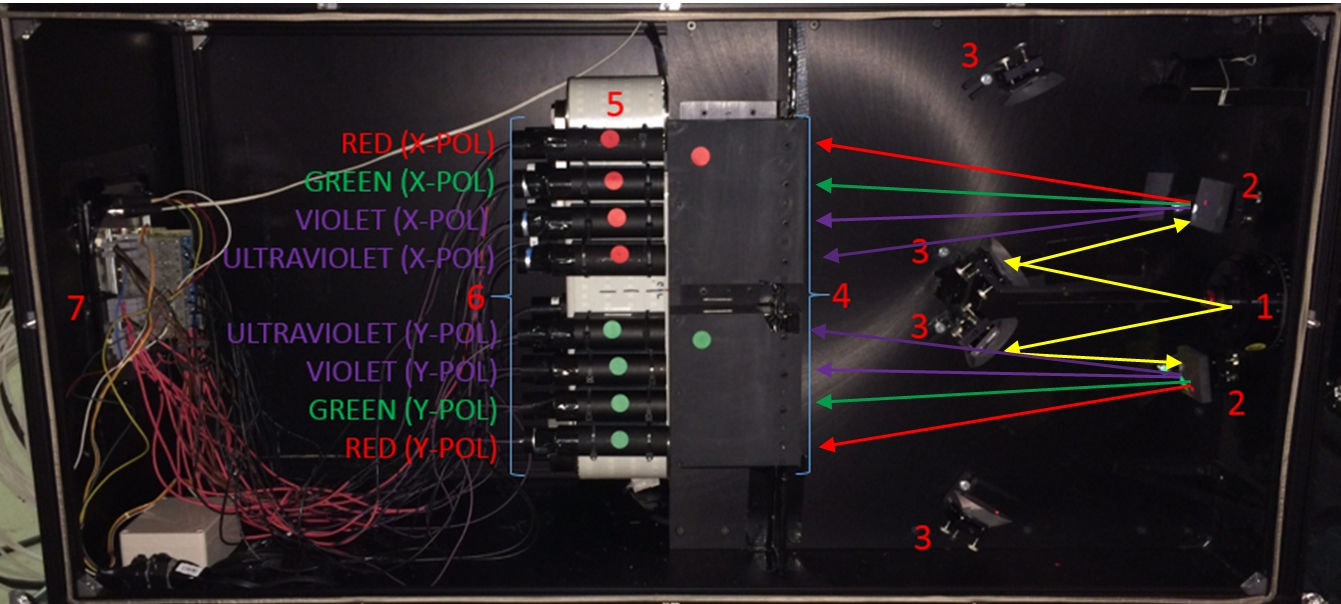} }}%
    \caption{(a) Vacuum mirrors and extraction windows used to extract the light inside the vacuum chamber. (b) The light extracted from the vacuum chamber travels through a LABM optical channel. (c) The light is driven outside the radiation area to a shielded area underneath through a manhole. (d) One side of an optical box used to measure the beamstrahlung. The elements contained in the box are: Wollaston prism (1), gratings (2), mirrors (3), lenses (4), a conveyor belt (5), photo-multipliers (6), and electronics (7).}%
    \label{fig:labm}%
\end{figure}

The first experimental observation of beamstrahlung took place at the Stanford Linear Collider (SLC)~\cite{PhysRevLett.62.2381}, colliding $e^{+}$ and $e^{-}$. The properties, polarization and spectrum, of the beamstrahlung are directly related to the beam parameters, and the analytical relations can be found in literature~\cite{Bassetti:1983ym,DiCarlo:2017hqj}. The LABM at SuperKEKB extracts visible beamstrahlung from about 5 meters distance downstream of the IP. The light is extracted by using vacuum mirrors inside the accelerator's vacuum chamber, as shown in Figure~\ref{fig:labm} (a), with the light then going through glass windows. The light is then driven through an optical channel with a series of mirrors and reaches an optical box after several meters, see Figure~\ref{fig:labm} (b, c). The mirrors are located inside a series of aluminum pipes connected to each other at 90 degrees. Since the beamstrahlung is emitted by both the electron and positron beams, the same apparatus is installed in both the electron ring, or High Energy Ring (HER), and the positron ring, or Low Energy Ring (LER). In each ring, there are two vacuum mirrors, located downstream of the IP, at the top and at the bottom of the vacuum chamber. This allows to be sensitive to vertical asymmetries with respect to the collision of the two beams. Since there are two vacuum mirrors on each ring, we have 4 optical channels, each one producing 8 PMTs measurements. Optical boxes, containing all the optical elements necessary to the measurements, are located outside the radiation area in order to minimize interference with the electronics. Each box is organized in two sides, each side accommodating 8 PMTs that serve one optical channel. Figure~\ref{fig:labm} (d) shows one side of one optical box, containing all the optical elements needed and the 8 PMTs located on the rear side of the box. There are 2 boxes and each box has two sides, providing therefore a total of 32 PMTs measurements. Each PMT receives light with horizontal or vertical polarization and in a different spectral region in the range going from about 390 nm to about 650 nm. Therefore, while there is some level of redundancy, in reality each PMT carries new information on the beam properties. 

\section{Data taking, data selection, observables }
\label{section:datas}
 In this first paper, we analyze only the data from the two electron telescopes, therefore 16 phototubes. These two are located at the top in the daisy chain of mirror motors. Regrettably, the use of a single data bus for all motors leads occasionally to interference that affects the higher numbered motors (those of the positron telescopes). Once this happens, the device can be reset only by access to the Interaction Region (IR). Still, with the positron telescopes not pointed at the IP we were able to cross check that there was no sensitivity on that side. We also pointed one of the telescopes in use to a feature that was clearly a reflection, with the same results.
 
 The results presented here, obtained through ML, solve a problem that has plagued this device for years. In an attempt to gain better knowledge of backgrounds, which were rapidly varying, the device was scanned around the IP spots presented in Fig.~3. In fact, this background subtraction method was plagued by other problems, such as rapidly varying backgrounds during beam refills (which are continuous at SuperKEKB), possibly varying beam tails in the quadrupoles, and other machine variations. The rapidity of such changes was generally larger of the ability of the device to move between side band and peak. Starting in 2018, our beamstrahlung signal got progressively noisier every year.
 Therefore, a classic "side band" subtraction scheme could not be successfully implemented.
 
 The use of opposite telescopes (one located at the top of the vacuum chamber, and one at the bottom) permits an automatic correction for the varying beam orbits which affect both signal (beamstrahlung) and background (synchrotron light from dipoles and quadrupoles).
 
\begin{figure}%
    \centering
    \includegraphics[width=\textwidth]{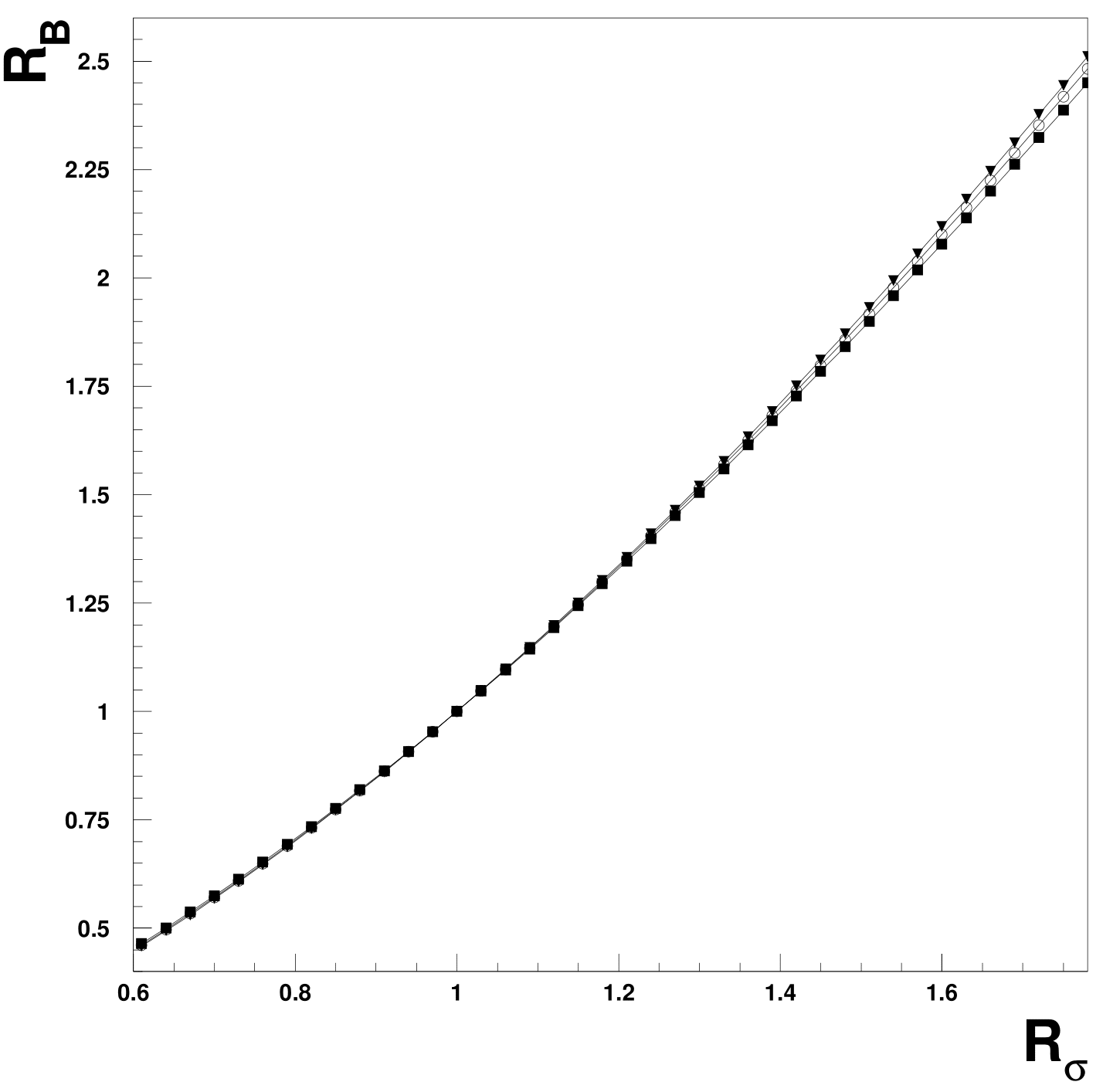}%
    \caption{Dependence of beamstrahlung rates on beam heights. In abscissa is the ratio of electron and positron beam height, $R_\sigma= \sigma_-/\sigma_+$. In ordinate is the observed $y-$polarized beamstrahlung ratio of rates, $R_B=R_-/R_+$. The calculation was performed using the SuperKEKB crossing angle, equally populated beams, with equal $\sigma_x$=10$\mu$m, and equal beam lengths $\sigma_z$=6mm. The three closely spaced lines, marked by different markers at the points of calculation, correspond, from lower to higher, to 3 different electron beam heights, respectively 60 nm, 160 nm, and 260 nm.}%
    \label{fig:sigyratio}%
\end{figure}

 Ref.~\cite{Bonvicini:1999rg} explains the observables used but a simple plot is provided, obtained from our unpublished large angle beamstrahlung calculations. These were performed using SuperKEKB beams, rigid beams approximation, at nominal and close to nominal beam parameters at the IP. The SuperKEKB beams are very flat, with an aspect ration $\sigma_y/\sigma_x$ at the IP of order 100. With beams this flat, the distance between particles, and therefore the deflecting force, is dominated by the horizontal distance. Fig.~\ref{fig:sigyratio} shows that the ratio of the two beam heights, can be deduced from the ratio of $y-$polarized beamstrahlung yields, on opposite sides, almost without dependence on the horizontal beam size. 

 The sensitivity to the ratio of beam heights expresses itself through both a reduction in $y-$polarized rate for the electron beam and an increase for the positron beam, when the positron beam size increases. Not having a two sided measurement, we resort to beam size variation to extract the same result during the course of our data taking.
 
 Specifically in the case of beamstrahlung a beam entering the IP with a small vertical angle $\delta$ will increase the rate of one telescope by a quantity of order $\delta/\theta$, and decrease the rate in the opposite telescope by $-\delta/\theta$, where $\theta$ is the angle of observation. This small correction is less or of order 1\% at SuperKEKB, and efficiently dealt with by the neural network. The neural network also can reproduce small effects such as photomultiplier saturation (a 1-3\% effect in these data), and effectively finds beamstrahlung, including spectral effects (when beams cross at an angle the spectrum at our observation angle, about 8 mrad, differs for x- and y-polarized light, see Ref.~\cite{DiCarlo:2017hqj}).
 
\begin{figure}%
    \centering
    \includegraphics[width=\textwidth]{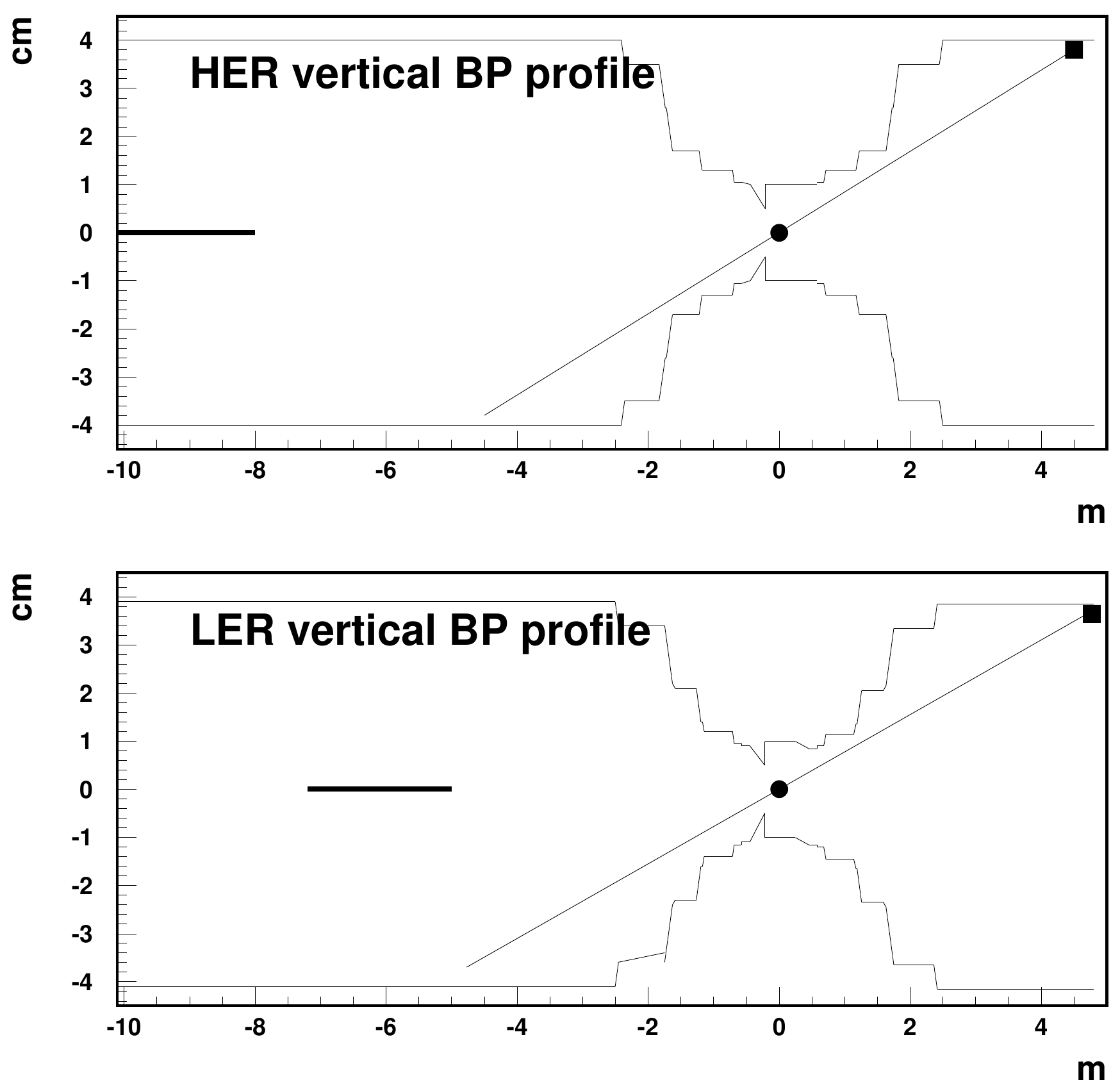}%
    \caption{Vacuum chamber, or Beam Pipe (BP), vertical profile at the SuperKEKB Interaction Point. Top: HER vertical vacuum chamber profile. Bottom: LER vertical vacuum chamber profile. The IP and the vacuum mirrors are shown to the right. The thick bars to the left are the locations and lengths of the last dipoles in the beam line.}%
    \label{fig:beampip}%
\end{figure}

 In order to analyze beamstrahlung, there needs to be certainty that the telescopes are pointed at the Interaction Point (IP). Fig.~\ref{fig:beampip} shows the vertical geometry for the electron telescope located below the vacuum chamber. The second mirror in the telescope can be oriented by two stepper motors to scan the field of view looking for light spots. 
 From Fig.~\ref{fig:beampip} the features of the light spot can be expected to be in the form of a lentil, with an angle equivalent to 1500 steps in the horizontal direction and 1500 in the vertical direction, or 0.7 mrad horizontal by 0.3 mrad vertical. The data presented here are taken with two 2 mm collimators in each telescope, separated by about 10 meters, providing a triangular acceptance with base 0.4 mrad in each direction.
 
\begin{figure}%
    \centering
    \includegraphics[width=\textwidth]{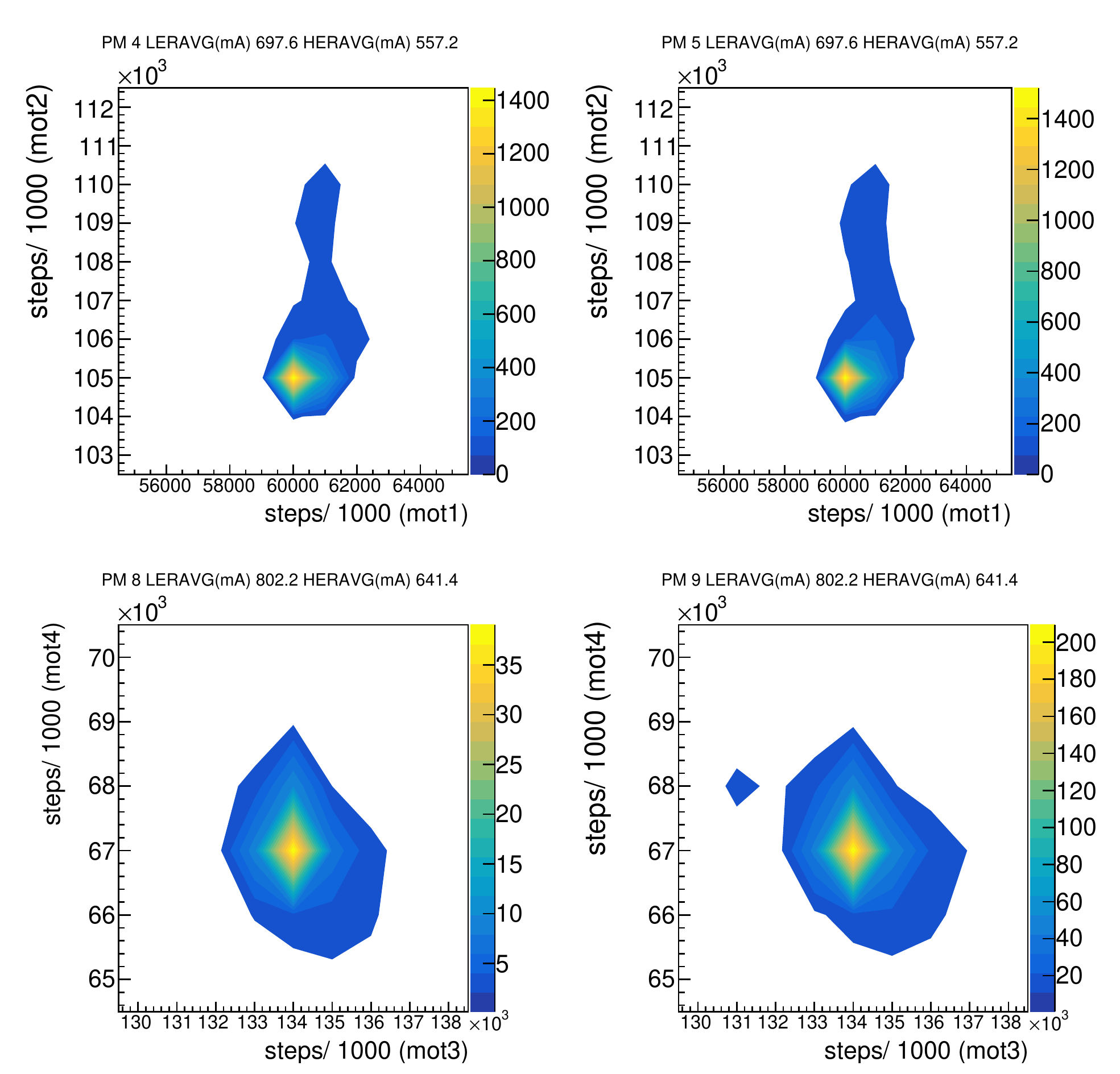} %
  
    \caption{Fine angular scans near the detected Interaction Point (IP). Top: electron (HER) telescope up. Bottom: electron (HER) telescope down. The IP is generally
    located very close to the maximum intensity spot.}%
    \label{fig:scans}%
\end{figure}

 Often telescopes, in this case the Down telescope, provide multiple spots of light. Past experience indicates that the IP spot must have correct dimension (specified above), and is generally more y-polarized (due to the presence of beamstrahlung. Reflected synchrotron light is often strongly x-polarized) and also generally can produce more elongated features compared to a proper spot. Figs.~\ref{fig:scans} show the angular scans for two PMTs for each telescope. We generally select the geometrical center of the spot as the IP.
 Although we have done much data taking in scanning mode (in the simplest case, by continuously measuring the presumed IP, and then two points at each horizontal side) in the past, looking to have "side band" measurements to measure independently signal and background, for the data presented here the motors were left at the IP without moving for 11 days. Subsequently, we took data at a spot considered fake in the down telescope, finding no correlation with beam parameters.
 
\begin{figure}%
    \centering
    \includegraphics[width=\textwidth]{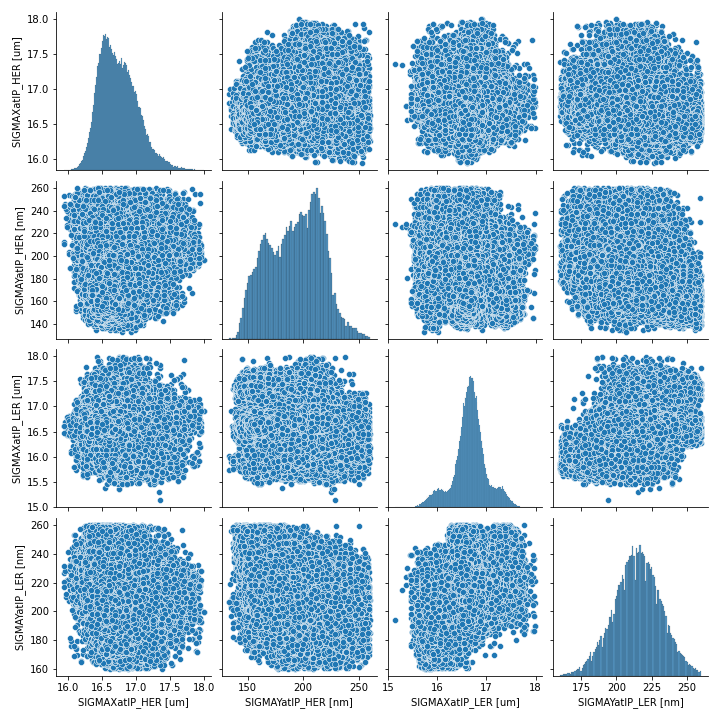}%
    \caption{Distributions of calculated beam parameters at the IP used in this paper, based on XRM \cite{Mulyani:2019gsy} beam size measurements at other locations in SuperKEKB. The plots along the diagonal represent the distributions for each parameter. The off-diagonal scatter plots show the distribution of any two pair of parameters. First
    line and first column: electron $\sigma_x$; second line and second column: electron $\sigma_y$; third line and third column: positron $\sigma_x$; fourth line and fourth column: positron $\sigma_y$.}%
    \label{fig:variation}%
\end{figure}

 Data used were subject to minimal cuts. We selected only Physics data, since any tuning can easily make our spots disappear. All data with currents above 100 mA were considered for analysis. Fig.~\ref{fig:variation} shows that during data taking the transverse sizes of both beams changed quite a bit, with little correlation between any pair of parameters, resulting in good parameter space coverage for our analysis.

\section{The Neural Network model}
\label{section:nn_model}
The model used in this work is a fully connected deep Neural Network (NN) with the following architecture in terms of neurons: 16-64-128-64-32-1. The model therefore consists of 19777 trainable parameters. The first layer consists of the 16 neurons of the input, i.e. the 16 PMT input values from the LABM, and the output layer consists of 1 neuron because we want to use our NN to perform a regression on one of the beam parameters. The hidden layers architecture was determined after several trial and error attempts in order to optimize the result of the regression. The model was implemented using Keras~\cite{chollet2015keras}, a high level abstraction library that works on top of the low level TensorFlow~\cite{tensorflow2015-whitepaper} compute engine. The data set used consists of about 150000 LABM and SuperKEKB measurements, which are split in about 96000 points for training, 24000 for validation, and 30000 for testing of the model, obtained over 11 days every five seconds. These measurements are randomly shuffled before being used, meaning that there is no time correlation between successive points in the data set. The about 30000 measurements reserved for testing do not take any part in the model learning process, allowing for an unbiased benchmark, and will be used to present the results in the next section, as they effectively represent new measurements with respect to the model.

\section{Reproduction of beam parameters}
\label{section:beam_parameters}
In this section we show the results of the NN model and we will also compare these results with those provided by the traditional Linear Regression (LR):
\begin{equation} \label{eq:linear_regression}  
y=\beta_{0}+\beta_{1}x_{1}+...+\beta_{16}x_{16}
\end{equation}

where the 16 independent variables $x_{i}$ correspond to the 16 PMT values provided by the current data selection. Because previously data were obtained by scanning, a direct comparison is not possible. A LR approach is feasible as, historically, we have always seen high linearity with beam current and other beam parameters in zero beamstrahlung conditions (that is, when only one beam was present in the accelerator).

The data used in this paper has been collected parasitically during a physics run for the Belle II experiment. In this sense, the beam parameters tend to be quite stable, but there are still significant changes that we can observe in the experimental data and try to reproduce with the predictions provided by our models. The goal of this study is to show that the variation of the beam parameters can be reproduced by a NN model with the LABM measurements as input. In the next subsections, we will compare experimental data and prediction by sorting them from smaller to larger values with respect to the experimental data. This kind of sorted plot is sometimes called a lift chart, and it is useful to evaluate the quality of a regression. Besides the visual evaluation, the value of the Mean Absolute Error (MAE) for each set of predictions is calculated. The relative MAE, defined as
\begin{equation} \label{eq:mae}  
\frac{1}{N}\sum_{i=1}^{N}\frac{|y_{i}-y_{i,pred}|}{y_{i}}
\end{equation}
where N is the number of measurements, $y_{i}$ is the measured value, and $y_{i,pred}$ the one predicted by the model, is indicated in the legend of each prediction plot as percentage error.

\subsection{Specific Luminosity}
\label{subsection:spec_lumi}
\begin{figure}%
    \centering
    \subfloat[]{{\includegraphics[width=5.5cm]{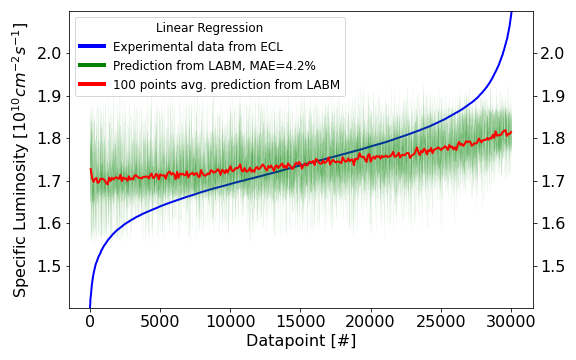} }}%
    \qquad
    \subfloat[]{{\includegraphics[width=5.5cm]{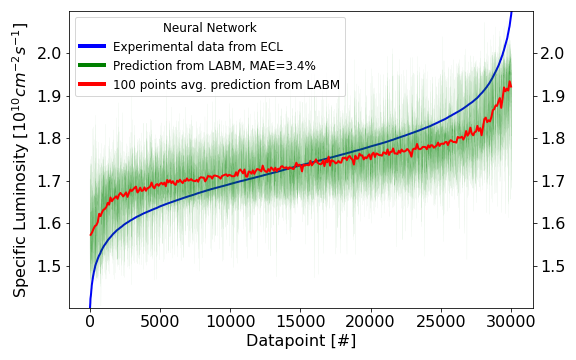} }}%
    \caption{(a) Specific Luminosity data and prediction with Linear Regression. (b) Specific Luminosity data and prediction with Neural Network.}%
    \label{fig:spec_luma}%
\end{figure}

\begin{figure}%
    \centering
    \subfloat[]{{\includegraphics[width=5.5cm]{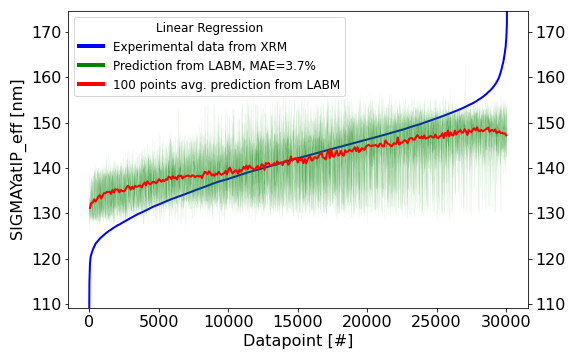} }}%
    \qquad
    \subfloat[]{{\includegraphics[width=5.5cm]{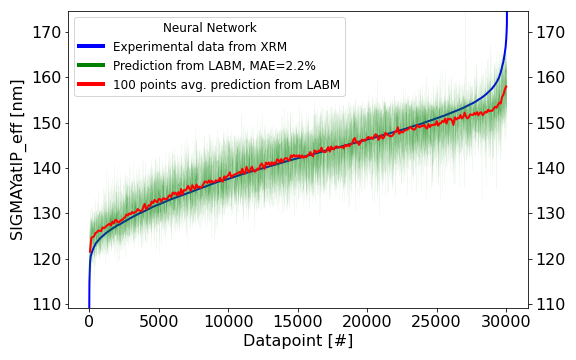} }}%
    \caption{(a) $\sigma_{y,eff}$ data and prediction with Linear Regression. (b) $\sigma_{y,eff}$ data and prediction with Neural Network.}%
    \label{fig:sigmaeff}%
\end{figure}

Although luminosity is not a beam parameter, it is strictly related to the beam parameters at the IP and it constitutes, together with the energy, one of the two figures of merit of a collider. Therefore, it will be the first experimental measurement that we will try to reproduce with our models. The absolute luminosity at SuperKEKB is measured by the Electromagnetic Calorimeter (ECL) monitor~\cite{Belle-ECL:2015vma}, located in the Belle II detector. The ECL measures Bhabha events and, following calibration, provides an absolute value for the luminosity. However, the luminosity depends on the currents and on the number of bunches present in the beam, and in our models we only want to use the 16 measurements from the electron side of the LABM as input. Therefore, for our purposes, we will use the specific luminosity. The specific luminosity for collinear Gaussian beams is defined as:
\begin{equation} \label{eq:spec_luminosity}  
L_{sp} = \frac{f_{0}}{2\,\pi\,\Sigma_{x}\,\Sigma_{y}}
\end{equation}
where $f_{0}$ (0.1MHz for SuperKEKB) is equal to the single bunch revolution frequency, and $\Sigma_{i}$ (i=x,y) are the convoluted beam sizes, corresponding to the quadrature sum of the two beam sizes at the IP: $\Sigma_{i}^{2}=(\sigma_{i,1}^{*})^{2}+(\sigma_{i,2}^{*})^{2}$. It is obtained from the regular formula of the luminosity dividing by the factor $N_{b}\,N_{1}\,N_{2}$, i.e., the number of bunches times the product of the numbers of particles per bunch of the two beams. In this way, the specific luminosity is independent from the beam currents and from the number of bunches present in the rings. Figure~\ref{fig:spec_luma} shows the specific luminosity as predicted by LR (a) and NN (b). From the comparison, we see that while the LR model reproduces the average value fairly well, it fails to predict the changes in specific luminosity at the low and high end of the plot. On the other hand, the NN is able to better predict and follow these changes.
Simple geometrical considerations show that in the SuperKEKB beam crossing situation the specific luminosity should closely track the variable
\begin{equation} \label{eq:sigma_eff}  
\sigma_{y,eff}=\frac{\sigma_{y,1}^{*}\sigma_{y,2}^{*}}{\Sigma_{y}}
\end{equation}
where $\sigma_{y,1}^{*}$ and $\sigma_{y,2}^{*}$ are the beam heights of the two beams at the IP and $\Sigma_{y}=\sqrt{(\sigma_{y,1}^{*})^{2}+(\sigma_{y,2}^{*})^{2}}$. In fact, SuperKEKB adopted the nano-beam scheme with a large crossing angle (83 mrad), and in such configuration the luminosity is approximately independent from the horizontal beam sizes. The tracking of this variable by the NN network is excellent, as shown in Figure~\ref{fig:sigmaeff}.
\subsection{Vertical beam size}
\label{subsection:beam_size}
\begin{figure}%
    \centering
    \subfloat[]{{\includegraphics[width=5.5cm]{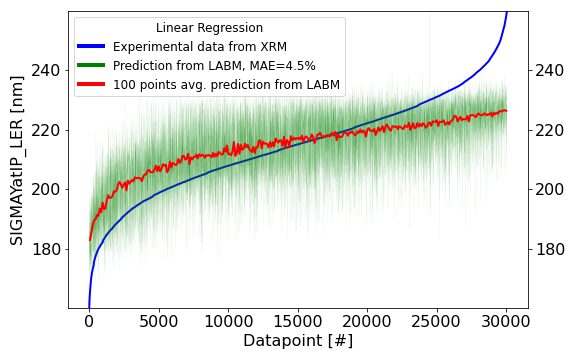} }}%
    \qquad
    \subfloat[]{{\includegraphics[width=5.5cm]{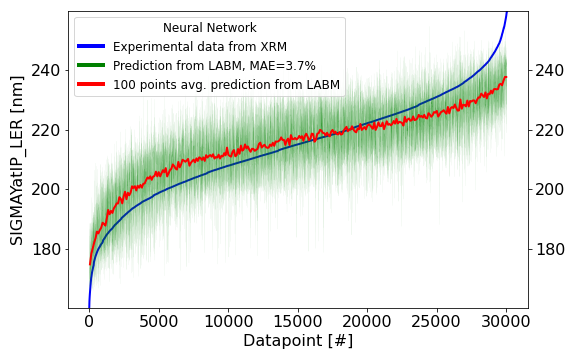} }}%
    \caption{(a) SIGMAY at IP for the LER as predicted with Linear Regression. (b) SIGMAY at IP for the LER as predicted with with Neural Network.}%
    \label{fig:sigmayl}%
\end{figure}

\begin{figure}%
    \centering
    \subfloat[]{{\includegraphics[width=5.5cm]{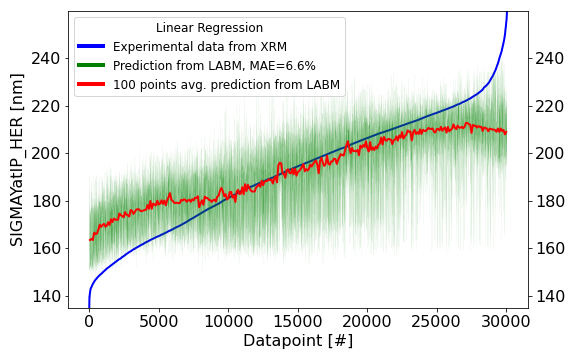} }}%
    \qquad
    \subfloat[]{{\includegraphics[width=5.5cm]{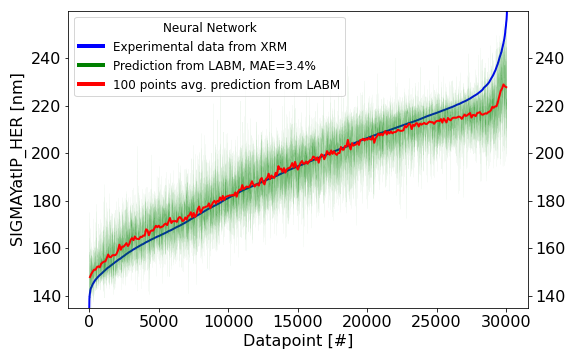} }}%
    \caption{(a) SIGMAY at IP for the HER as predicted with Linear Regression. (b) SIGMAY at IP for the HER as predicted with a Neural Network.}%
    \label{fig:sigmayh}%
\end{figure}

\begin{figure}%
    \centering
    \subfloat[]{{\includegraphics[width=5.5cm]{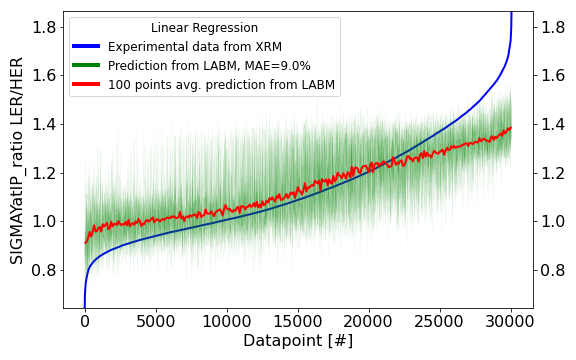} }}%
    \qquad
    \subfloat[]{{\includegraphics[width=5.5cm]{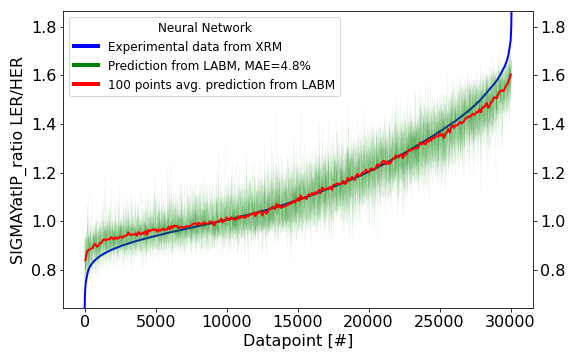} }}%
    \caption{(a) Ratio SIGMA Y LER/HER data and prediction with Linear Regression. (b) Ratio SIGMA Y LER/HER data and prediction with Neural Network.}%
    \label{fig:sigmayr}%
\end{figure}

\begin{figure}%
    \centering
    \subfloat[]{{\includegraphics[width=5.5cm]{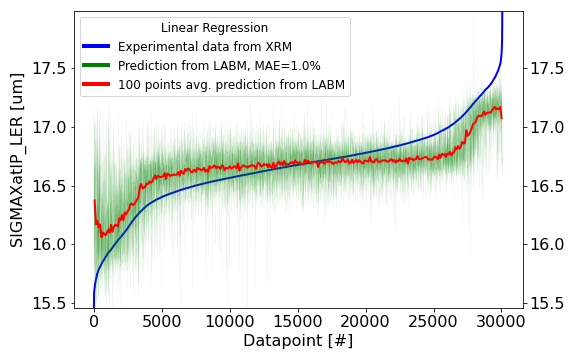} }}%
    \qquad
    \subfloat[]{{\includegraphics[width=5.5cm]{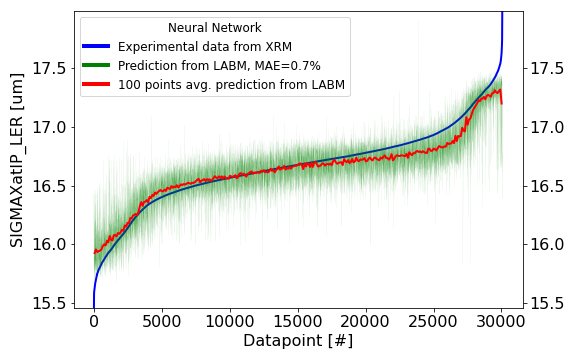} }}%
    \caption{(a) SIGMAX at IP for the LER as predicted with Linear Regression. (b) SIGMAX at IP for the LER as predicted with with Neural Network.}%
    \label{fig:sigmaxl}%
\end{figure}

\begin{figure}%
    \centering
    \subfloat[]{{\includegraphics[width=5.5cm]{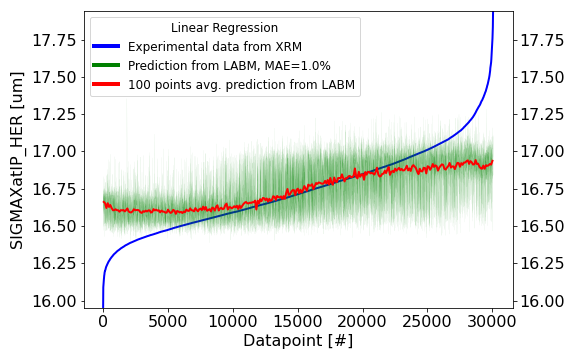} }}%
    \qquad
    \subfloat[]{{\includegraphics[width=5.5cm]{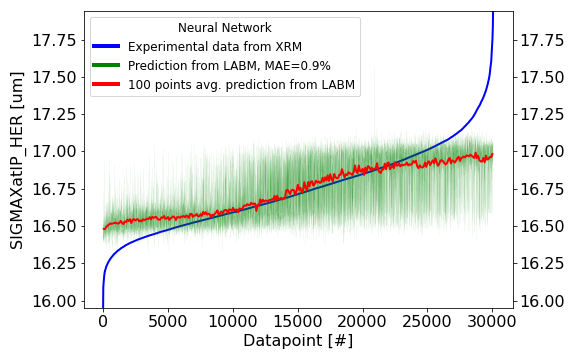} }}%
    \caption{(a) SIGMAX at IP for the HER as predicted with Linear Regression. (b) SIGMAX at IP for the HER as predicted with a Neural Network.}%
    \label{fig:sigmaxh}%
\end{figure}

\begin{figure}%
    \centering
    \subfloat[]{{\includegraphics[width=5.5cm]{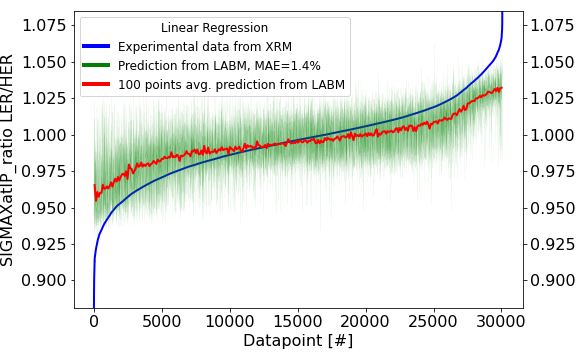} }}%
    \qquad
    \subfloat[]{{\includegraphics[width=5.5cm]{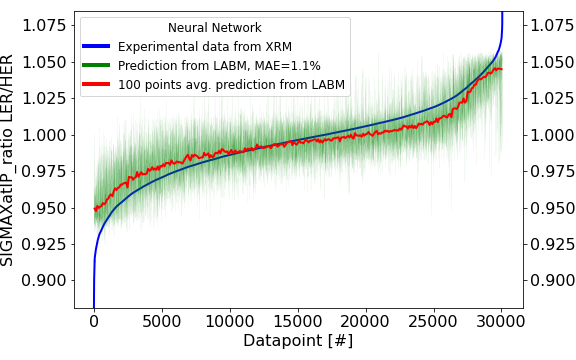} }}%
    \caption{(a) Ratio SIGMA X LER/HER data and prediction with Linear Regression. (b) Ratio SIGMA X LER/HER data and prediction with Neural Network.}%
    \label{fig:sigmaxr}%
\end{figure}

The beam size at SuperKEKB is measured by the X-ray monitor (XRM)~\cite{Mulyani:2019gsy}. The XRMs are installed in both the HER $(e^{-})$ and LER $(e^{+})$ rings, at 641.4 and 1397.7 meters respectively from the IP. Using the Twiss parameters at their location, it is possible to obtain an estimate of the emittance, and through the optical transfer matrices it is possible to estimate the beam size at the IP, which is the quantity we are interested in. Figure~\ref{fig:sigmayl} shows data and prediction of SIGMAY at IP for the LER ring. In this case the predictive power of LR and NN is similar, although we can appreciate how the NN is able to better reproduce the variation in beam size for the highest values.
In the case of SIGMAY at IP for the HER beam, we see a much better predictive power for the NN. Figure~\ref{fig:sigmayh} shows that the NN prediction is much less noisy and the MAE is 6.6\% for the LR and 3.4\% for the NN.
Finally, we are interested in the LER/HER ratio on SIGMA Y at IP. In fact, KEKB had a vertical beam size for the LER that was consistently larger than the corresponding one for the HER. This corresponds to one beam being unfocused, causing significant luminosity degradation, which is an effect that we want to prevent at SuperKEKB. Figure~\ref{fig:sigmayr} shows the ratio sigmay LER/HER, showing as well that the NN model predicts the experimental data much better than the LR, which is very noisy, with the MAE being 9.0\% for LR and 4.8\% for the NN.
We did not focus on the horizontal beam sizes since they are very stable and of little interest with respect to SuperKEKB's luminosity, as discussed above in Section \ref{subsection:spec_lumi}. However, for completeness the plots related to horizontal beam widths and their ratio are shown in Figs.~\ref{fig:sigmaxl} to ~\ref{fig:sigmaxr}. The good accuracy in predicting the horizontal sizes and ratio is noted, also with errors of order percent.
\section{Discussion}
\label{section:discussion}
In order to understand the results of this paper, it is worth summarizing what a single side observation can and cannot do. The technique was first proposed as a two-side observation ~\cite{Bonvicini:1999rg}. The two-side
analysis would then be able to reconstruct beam parameters. One of the main results of this paper is that single side measurements have good sensitivity to all four parameters. A two-side measurement next year will then be able to further reduce the error, while also provide auxiliary measurements of the beam tails. With two side observation the LABM beam parameters determination will
become truly independent of other sensors. In this paper we limit ourselves to correlating LABM parameters with another, existing sensor.

The beams are very flat (meaning that $\sigma_x\gg \sigma_y$), and the NN did not know our signal rate calculations with the parameters of Table ~1. 
In such a situation, the ratio of beam heights can and should be measured well, but there is minimal sensitivity to each beam height. Nevertheless it is clear from Figs.~\ref{fig:sigmayh} and~\ref{fig:sigmayr} that the NN is able to predict well also the height of the electron beam (and, having the ratio and one size, the other size is derived also, Fig.~\ref{fig:sigmayl}). This necessarily entails the presence of another source of radiation which depends on the radiating beam height, which we identify as the light emitted by the vertical beam tails in the final quadrupoles. We can arrive at this conclusion by exclusion (dipole sources do not depend on the beam height).
It appears that the NN is able to disentangle the beamstrahlung and quadrupole radiation, a task that is currently beyond our abilities using standard analysis techniques.

The size of the electromagnetic fields of a beam scales like $1/\sigma_x\sigma_z$ for flat beams. A positron beam with a smaller horizontal size will make the electron beam radiate more, and vice versa. Therefore, without knowing the normalization, only the ratio of the beams horizontal sizes should be available through a beamstrahlung measurement. However, the same radiation from the final quadrupoles is present and can be used by the NN to provide an independent measurement.
Note that the hypothesized quadrupole radiation is not necessarily better or worse at measuring the horizontal size compared to the vertical size. 
The horizontal size is much larger than the vertical size, leading to a larger average displacement of particles from the center of the quadrupole, and therefore there will generally be more x-polarized radiation.
However, the vacuum mirrors are located vertically from the beam line axis, and particles bent vertically will illuminate them at a smaller (or larger) angle compared to particles bent horizontally. From the general
observed rates (with ratios between x-polarized and y-polarized radiation of order 2) it appears that this enhancement make the two polarized observed rates comparable.

It is noted that the NN provided measurements of beam parameters at the few percent level, which is crucial for the viability of the device as a beam monitor. Table~\ref{table:2} summarizes the relative MAE on the LR and NN predictions presented in the previous section, representing the main result of this work. 
\begin{table}[ht]
\centering
\begin{center}
\begin{tabular}{|c|c|c|}
\hline
$ Model $        & LR & NN \\ \hline
$L_{sp} $      & 4.2\%  & 3.4\%     \\ \hline
$\sigma_{y,eff} $  & 3.7\% & 2.2\%      \\ \hline
$\sigma_{x,LER}$ & 1.0\% & 0.7\%      \\ \hline
$\sigma_{x,HER}$ & 1.0\% & 0.9\%      \\ \hline
$\sigma_{x,LER}/\sigma_{x,HER}$ & 1.4\% & 1.1\%  \\ \hline
$\sigma_{y,LER}$ & 4.5\% & 3.7\%     \\ \hline
$\sigma_{y,HER}$ & 6.6\% & 3.4\%      \\ \hline
$\sigma_{y,LER}/\sigma_{y,HER}$ & 9.0\% & 4.8\%\\ \hline
\end{tabular}
\end{center}
\caption{Summary of relative MAE for LR and NN models for the results presented in this paper. The NN performs consistently better than the LR with errors at a few percent level.}
\label{table:2}
\end{table}
We also note that the main current NN limitation is just due to scarce statistics at the edges of the measured distributions, i.e. in the regions of low or high luminosity and low or high beam sizes. We stress that improvement is expected over time, allowing for larger and lower values of the parameter space contained in the measured data set to become more populated, therefore increasing the performance of the NN training, and consequently the quality of the predictions in those regions.

\section{Conclusion}
\label{section:conclusion}

This study had two main purposes: (1) to show that the LABM measurements of beamstrahlung signal are of excellent quality and are correlated with key beam parameters; (2) to show that machine learning, in this case in the form of a Neural Network model, can be useful in modern particle accelerators with extremely small beams and large sensitivities to beam parameters. In section~\ref{section:beam_parameters} we showed that the NN model was excellent at predicting specific luminosity and beam sizes for unseen data using as input only 16 of the 32 PMT values from the LABM. The data was collected parasitically during a Physics Run, and therefore large variation in the parameters is not to be expected. However, the NN model was especially good at reproducing the modest changes, performing always better or much better compared to the LR model. The NN models, one for each of the parameters reproduced, can in principle be deployed online to provide an estimation for the given parameters by taking input from the 16 PMT values. This could also be useful when one or more of the other instruments are offline, as in this case the LABM can provide a temporary replacement value. The ability to predict specific luminosity and beam size from the LABM measurements only was at the same time an independent validation of the LABM quality itself. The next step for the LABM would be to introduce functions of some or all of the 32 PMT values, based on theoretical ground or machine learning techniques, to provide new and original information about the beam parameters at the IP. Finally, this study shows that machine learning techniques can be useful in modern accelerators. SuperKEKB's instrumentation provides hundreds of variables other than LABM's that could in principle be used in larger correlation studies, significantly increasing the predictive power.
\par
This study has shown the application of machine learning techniques in the effort of reproducing key beam parameters at SuperKEKB. The Neural Network model used had a significantly larger predictive power compared to the classical Linear Regression. The model used only the LABM measurements as input, and was able to predict specific luminosity and vertical beam sizes. This constituted on one hand a validation of the LABM itself, and on the other hand a further validation of machine learning techniques use in accelerator physics.

\section*{Acknowledgments}
We thank David Cinabro for useful discussions.
This work was supported by the Frontiers of Science Program Contracts
No. FOINS-296, No. CB-221329, No. CB-236394,
No. CB-254409, No CB-A1-S-33202 and No. CB-180023, Profapi PRO-A1-018, SEP-CINVESTAV research Grant No. 237, University of Tabuk, KSA,  research grant S-0265-1438, and by the U.S. Department of Energy, Office of Science, through grant SC-007983 and the US-Japan Science and Technology
Cooperation Program. We thank the beam instrumentation, commissioning, and vacuum groups of the SuperKEKB accelerator. 


\bibliography{mybibfile}

\end{document}